\RequirePackage{lineno}
\documentclass[prd,twocolumn,nofootinbib,superscriptaddress,floatfix]{revtex4}
\usepackage{cancel}
\usepackage{graphicx}
\usepackage{epsfig}
\usepackage{amsmath,amsfonts,amssymb}
\usepackage{t1enc}

\newcommand{\mnmw}{(M_{W_R},m_N)}

\newcommand{\etmiss}{E_T^\text{miss}}

\begin{document}


\title{Measuring heavy neutrino couplings at the LHC}

\author{J. A. Aguilar-Saavedra}
\affiliation{Departamento de F\'{\i}sica Te\'orica y del Cosmos, Universidad de Granada,
 Granada, Spain}
 \affiliation{Instituto de F\'{\i}sica de Cantabria (CSIC-UC), Santander, Spain}
\author{F. R. Joaquim}
\affiliation{Departamento de F\'{\i}sica and CFTP, Instituto Superior T\'ecnico, Universidade T\'ecnica de Lisboa, Lisboa, Portugal}


\begin{abstract}
The existence of heavy neutrinos mediating neutrino masses via a type-I seesaw can be directly probed at the LHC, or indirectly in rare lepton flavor-violating processes. The synergy between these two approaches requires a direct measurement of the heavy neutrino couplings. We discuss a strategy to perform such measurements at the LHC in the context of left-right symmetric models, which is also applicable to other models implementing a type-I or type-III seesaw. We demonstrate that the ambiguities in the determination of the heavy neutrino mixing parameters can be resolved by performing an exclusive analysis of dilepton final states, discriminated by flavor and missing energy.
\end{abstract}

\maketitle

\section{Introduction}

The smallness of light neutrino masses, compared to the other fermions, suggests that new physics beyond the standard model (SM) is involved in their generation. The simplest and most popular explanation for neutrino mass suppression is provided by the seesaw mechanism~\cite{Minkowski:1977sc}, which introduces heavy right-handed neutrinos. These new states need not to be at the TeV scale, nor they are necessarily at the direct reach of the Large Hadron Collider (LHC). Nevertheless, this is an interesting possibility which has motivated much effort both from the theoretical~\cite{Dicus:1991wj,Datta:1993nm,Han:2006ip,delAguila:2007em,delAguila:2008cj,delAguila:2008hw,Das:2012ze,delAguila:2007ua,Huitu:2008gf,AguilarSaavedra:2009ik,Li:2010rb,Keung:1983uu,Ferrari:2000sp,Gninenko:2006br,Nemevsek:2011hz,AguilarSaavedra:2012fu,Das:2012ii} and experimental~\cite{CMSN,atlas} sides.

It is well known that the heavy neutrinos may enhance the rates of lepton flavor-violating processes such as $\mu \to e \gamma$, $Z \to e\mu$, $\mu \to eee$ and $\mu-e$ conversion (see~\cite{Abada:2007ux,Branco:2011zb} for reviews). In this sense, precision measurements in the intensity frontier constitute a complementary window to testing the seesaw mechanism. Notably, the interplay between the high-energy and high-intensity probes of the seesaw requires the direct measurement of the heavy neutrino couplings at the LHC.
In this work we investigate how these couplings could be measured in case a heavy neutrino signal is observed.

As a benchmark scenario, we consider heavy neutrino production in the context of left-right symmetric models~\cite{Pati:1974yy}, for which the LHC reach is very good~\cite{Keung:1983uu,Ferrari:2000sp,Gninenko:2006br,Nemevsek:2011hz,AguilarSaavedra:2012fu,Das:2012ii}. In these scenarios, heavy neutrinos can be produced from the decay of an SU(2)$_R$ gauge boson $W_R$,
\begin{equation}
pp \to W_R \to \ell N \,,
\label{ec:prod}
\end{equation}
and they subsequently decay through an off-shell $W_R$ as
\begin{equation}
N \to \ell' jj\,,~\ell' tb \,,
\label{ec:dec}
\end{equation}
giving a $\ell \ell' jj$ signal, where $\ell,\ell' = e,\mu,\tau$ (with the same or opposite charges, due to the Majorana nature of $N$) and $j$ are light jets. Until now, almost all studies~\cite{Keung:1983uu,Ferrari:2000sp,Gninenko:2006br,Nemevsek:2011hz,Das:2012ii} dedicated to the study of heavy neutrino signals in left-right models have been concentrated on the simplest case where $\ell,\ell' = e,\mu$, ignoring the production of $\tau$ leptons. However, current experimental searches are sensitive to the $\tau$'s~\cite{AguilarSaavedra:2012fu} through the detection of secondary leptons produced in the leptonic decays $\tau \to e/\mu \nu \bar \nu$. We stress that, from the theoretical point of view, there is no compelling argument in favor of zero $\tau N$ mixing. Actually, bearing in mind the lepton mixing pattern revealed by neutrino oscillation experiments, one could easily find a motivation against that assumption. In view of this, an analysis including the $\tau$-induced dilepton final states is most welcome.

In this paper we will assume that a $\ell \ell' jj$ heavy neutrino signal is seen at the LHC. After determining the $W_R$ and heavy neutrino masses, which is quite straightforward from the observed resonance peaks, the next step would be to measure the couplings of the heavy neutrino to the charged leptons and the $W_R$ boson. This interaction is parametrized by the Lagrangian
\begin{equation}
\mathcal{L} = -\frac{g_R}{\sqrt 2} V_{\ell N}^R \bar \ell_R \gamma^\mu N_{R} W_{R\mu}^{-} + \text{H.c.} \,,
\end{equation}
where $g_R$ and $V_{\ell N}^R$ are the $\text{SU}(2)_R$ gauge coupling and $\ell N$ mixing parameters, respectively. The latter obey the general normalization condition
\begin{equation}
 |V_{e N}^R|^2 +  |V_{\mu N}^R|^2 + |V_{\tau N}^R|^2 \equiv \kappa^2  \,,
 \label{ec:kappa}
 \end{equation}
with $\kappa$ constrained to be very close to unity. For simplicity, in this work we consider that only one heavy neutrino $N$ is lighter than the $W_R$ boson. In case there is more than one, the method presented here can be extended by applying suitable kinematical cuts to single out each heavy neutrino contribution.

\section{Measuring heavy neutrino mixings in dilepton events}

Let us now discuss in some detail how the mixings can be extracted from event counting in dilepton final states, without assuming unitarity of the leptonic mixings, i.e. $\kappa \neq 1$. The cross section for the process in Eq.~(\ref{ec:prod}) with a given lepton $\ell$ is
\begin{equation}
\sigma(\ell N) = \sigma_{W_R} \frac{|V_{\ell N}^R|^2}{\kappa^2 + R} \,,
\end{equation}
where $\sigma_{W_R}$ is the cross section for $W_R$ production and
\begin{equation}
R = \left. \frac{\Gamma_\text{had}}{\Gamma_\text{lep}} \right|_{\kappa=1}\,,
\end{equation}
is the ratio between the hadronic and leptonic $W_R$ decay widths for $\kappa=1$ (this quantity is independent of lepton mixings).
The total cross section for $N$ production is then
\begin{equation}
\sigma_\text{tot} = \sigma_{W_R} \frac{\kappa^2}{\kappa^2 + R} \,,
\label{ec:sigmatot}
\end{equation}
which, together with the branching ratio for the decay in Eq.~(\ref{ec:dec}),
\begin{equation}
\text{Br}(N \to \ell' X) = \frac{ |V_{\ell' N}^R|^2}{\kappa^2} \,,
\end{equation}
leads to the cross section for the production of charged leptons $\ell$, $\ell'$,
\begin{equation}
\sigma_{\ell \ell'} = \frac{\sigma_\text{tot} }{\kappa^4} |V_{\ell N}^R|^2 |V_{\ell' N}^R|^2 \,.
\label{ec:sigmall}
\end{equation}
By introducing the normalized mixings
\begin{equation}
v_{\ell N}^R = \frac{V_{\ell N}^R}{\kappa}\,,
\end{equation}
we find the expected result
\begin{equation}
\sigma_{\ell \ell'} = \sigma_\text{tot} |v_{\ell N}^R|^2 |v_{\ell' N}^R|^2 \,,
\label{ec:sigmall2}
\end{equation}
i.e., the total heavy neutrino cross section is distributed among the different dilepton flavors according to the squares of the normalized mixings. The above equation also reflects the fact that one cannot simultaneously measure the total cross section and test the unitarity of the mixings from $\sigma_{\ell \ell'}$ alone. However, if the gauge coupling $g_R$ could be determined from other processes (for example, in the decay channel $W_R \to t\bar b$) and a definite prediction for $\sigma_{W_R}$ and $R$ could be made, then the unitarity of leptonic mixing would be testable by inverting Eq.~(\ref{ec:sigmatot}),
\begin{equation}
k = \left( \frac{R \, \sigma_\text{tot}/\sigma_{W_R}}{1-\sigma_\text{tot}/\sigma_{W_R}} \right)^{1/2} \,.
\label{ec:kappa2}
\end{equation}

We now address the question on how the individual cross sections $\sigma_{\ell \ell'}$ (and, thus, the mixing parameters) can be extracted from the data. We restrict ourselves to final-state signatures with two electrons or muons, which can result from prompt production $\ell,\ell' = e,\mu$ or from the leptonic decay of $\ell,\ell' = \tau$. Notice that with the tagging of hadronic $\tau$ decays the final state of one electron or muon plus a $\tau$-jet could be used as well, improving the results presented here. In a previous work~\cite{AguilarSaavedra:2012fu}, it has been emphasized that the key for the sensitivity to $\ell,\ell' = \tau$ in the dilepton final state is to drop any requirement on the missing energy of the events, $\etmiss$, which is small (and caused by instrumental effects) for $\ell,\ell' = e,\mu$ but large when there is one or more $\tau$ leptons. However, this is not enough to obtain precise measurements of the mixings and, therefore, a separation of the dilepton samples according to $\etmiss$ is necessary. This fact, which will be confirmed later by our numerical results, can be easily understood with the following simple semi-quantitative argument. Let us assume the same efficiency $\varepsilon$ for electron and muon detection. At high transverse momenta, as it corresponds to the production of a heavy $W_R$, the efficiency for secondary leptons originated from $\tau$ decays is similar~\cite{AguilarSaavedra:2012fu}, but these are penalized by the branching ratio $\text{Br}(\tau \to e \nu \bar \nu) \simeq \text{Br}(\tau \to \mu \nu \bar \nu) \simeq 0.17 \equiv \eta$. Then, for an integrated luminosity $L$ the number of events in each dilepton channel is
\begin{eqnarray}
N_{ee} & = & \sigma_\text{tot} L \varepsilon \left( |v_{eN}^R|^4 + 2 \eta |v_{eN}^R|^2 |v_{\tau N}^R|^2 + \eta^2 |v_{\tau N}^R|^4  \right) \,, \notag \\
N_{e\mu} & = & 2\sigma_\text{tot} L \varepsilon \,  \left[ |v_{eN}^R|^2 |v_{\mu N}^R|^2 \right. \notag \\
& & \left. + \eta \left(
|v_{eN}^R|^2 + |v_{\mu N}^R|^2 \right) |v_{\tau N}^R|^2 + \eta^2 |v_{\tau N}^R|^4
\right] \,, \notag \\
N_{\mu\mu} & = & \sigma_\text{tot} L \varepsilon \left( |v_{\mu N}^R|^4 + 2 \eta |v_{\mu N}^R|^2 |v_{\tau N}^R|^2 + \eta^2 |v_{\tau N}^R|^4  \right) \,. \notag \\
\label{ec:deg}
\end{eqnarray}
Although there are three independent parameters (for example, $v_{eN}^R$, $v_{\mu N}^R$ and $\sigma_\text{tot}$) and three measurements, it is clear that very distinct parameter space points can give roughly the same numbers of events in the three channels.\footnote{This degeneracy is not found in~\cite{Das:2012ii}, because $V_{\tau N}^R = 0$ is implicitly assumed.} In particular, along the line $v_{eN}^R = v_{\mu N}^R$ one has $N_{ee} = N_{\mu \mu} = 1/2 \, N_{e\mu}$, and the system of equations is underconstrained. For instance, the cases of maximal $eN$ and $\mu N$ mixing with $\sigma_\text{tot} \equiv \sigma_0$
\begin{equation}
v_{eN}^R = v_{\mu N}^R = \frac{1}{\sqrt 2} \,, \quad v_{\tau N}^R = 0 \,,\quad \sigma_\text{tot} \equiv \sigma_0\,,
\end{equation}
and the one with $\tau$ mixing only and $\sigma_\text{tot} \equiv \sigma_0 / \eta^2$,
\begin{equation}
v_{eN}^R = v_{\mu N}^R = 0 \,, \quad v_{\tau N}^R = 1 \,,\quad \sigma_\text{tot} \equiv \sigma_0 / \eta^2 \,,
\end{equation}
lead to the same number of events. The obvious difference between these two cases is that, in the latter, the events have large missing energy due to the neutrinos produced in $\tau$ decays. Then, the solution to the degeneracy is to separate the dilepton samples in `low $\etmiss$' and `high $\etmiss$' events, according to some convenient threshold. This distinction doubles the number of measurements and definitely resolves the aforementioned ambiguities, which are otherwise present when event counting is performed on the $\etmiss$-inclusive samples.

\section{Numerical analysis}

In order to test the above method in the extraction of heavy neutrino mixings, we rely on a fast simulation analysis of heavy neutrino signals and corresponding SM backgrounds at the LHC, with a center of mass energy of 14 TeV. A luminosity of 30 fb$^{-1}$ is assumed. Signal events are generated with {\sc Triada}~\cite{AguilarSaavedra:2009ik} and backgrounds with {\sc Alpgen}~\cite{Mangano:2002ea}. The hadronization is performed by {\sc Pythia}~\cite{Sjostrand:2006za}, and {\sc AcerDet}~\cite{RichterWas:2002ch} is used for a fast simulation of a generic LHC detector. For the suppression of the SM backgrounds we apply an event selection similar to the one used by the ATLAS Collaboration for 7 TeV searches~\cite{atlas}, but tightening some of the kinematical cuts. Namely, we require:
\begin{itemize}
\item Exactly two leptons with transverse momentum $p_T > 50$ GeV, and pseudo-rapidity $|\eta| < 2.47$ for electrons, $|\eta| < 2.4$ for muons. Electrons in the range $1.37 < |\eta| < 1.52$ are excluded.
\item At least one jet with $p_T > 20$ GeV and $|\eta| < 2.8$.
\item Dilepton invariant mass $m_{\ell \ell} > 110$ GeV.
\item The sum of transverse momenta of the two leptons and highest $p_T$ jets (including up to two jets) must be larger than 700 GeV.
\item The $W_R$ reconstructed mass must be larger than 700 GeV. For events with at least two jets, this mass is defined as the invariant mass of the two leptons and the two leading jets, $M_{W_R}^\text{rec} = m_{\ell \ell jj}$. For events with only one jet, it includes only this jet, $M_{W_R}^\text{rec} = m_{\ell \ell j}$.
\end{itemize}
We split our dilepton final state events in twelve disjoint sets. First, those with same-sign leptons are separated from the ones with opposite-sign. This turns out to be convenient since the former have a much lower SM background. Second, each one of those two sets is divided in two subsets with $\etmiss < 200$ GeV and $\etmiss > 200$ GeV, corresponding to `low' and `high' $\etmiss$. Finally, the events are classified according to the lepton flavour. The SM background $B_X$ for each final state generically labelled as $X$ is given in Table~\ref{tab:bkg} for a luminosity $L = 30~\text{fb}^{-1}$. It includes $t \bar t$, single top, $W/Z\,$ $+$ jets, $W/Z\, b \bar b$ $+$ jets, $W/Z\, c \bar c$ $+$ jets,
$W/Z\, t \bar t$ $+$ jets, diboson and triboson production. Further details about the background generation can be found in~\cite{AguilarSaavedra:2009ik}.

\begin{table}[htb]
\caption{SM background $B_X$ for the twelve dilepton final states $X$ and for a luminosity of 30 fb$^{-1}$
\label{tab:bkg}}
\begin{center}
\begin{tabular}{cccccccccccc}
\hline
\hline
\multicolumn{6}{c}{Same sign} & \multicolumn{6}{c}{Opposite sign} \\
\multicolumn{3}{c}{Low $E_T^\text{miss}$} & \multicolumn{3}{c}{High $E_T^\text{miss}$}
& \multicolumn{3}{c}{Low $E_T^\text{miss}$} & \multicolumn{3}{c}{High $E_T^\text{miss}$} \\
 $ee$ & $e\mu$ & $\mu\mu$ & $ee$ & $e\mu$ & $\mu\mu$ & $ee$ & $e\mu$ & $\mu\mu$ & $ee$ & $e\mu$ & $\mu\mu$ \\
11 & 15 & 8 & 3 & 3 & 2 &
1180 & 702 & 1316 & 91 & 203 & 227
\\
\hline
\hline
\end{tabular}
\end{center}
\end{table}

For the heavy neutrino signal we assume $g_R = g_L$ and consider two mass points: (1) $M_{W_R} = 2.5$ TeV, $m_N = 1$ TeV, for which $\sigma_{W_R} = 490$ fb, $R = 11.8$, $\sigma_\text{tot} = 38.3$ fb; (2) $M_{W_R} = 2$ TeV, $m_N = 1$ TeV, for which $\sigma_{W_R} = 1580$ fb, $R = 14.2$, $\sigma_\text{tot} = 104$ fb. For each case we generate nine signal samples with $\ell,\ell' = e,\mu,\tau$. These samples allow us to obtain the signal resulting from a heavy neutrino with arbitrary mixing, by an appropriate averaging with weights $w_{\ell \ell'} = |v_{\ell N}^R|^2 |v_{\ell' N}^R|^2$. (Note that $\sum w_{\ell \ell'} = 1$ because $\sum |v_{\ell N}^R|^2 = 1$.) The efficiencies $\varepsilon_X^{\ell \ell'}$ after cuts for the nine samples are given in Tables~\ref{tab:eff1} and \ref{tab:eff2} for the mass points 1 and 2, respectively. Notice that the $\varepsilon_X^{\ell \ell'}$ are rather similar for the two cases, with some migration between the low $\etmiss$ and high $\etmiss$ final states due to the different kinematics. Then,
the number of signal events for a given final state $X$ is
\begin{equation}
S_{X} = \sigma_\text{tot} L \sum_{\ell \ell'} w_{\ell \ell'} \varepsilon_X^{\ell \ell'} \,.
\end{equation}
with $\sigma_\text{tot}$ given by Eq.~(\ref{ec:sigmatot}).
\begin{table}[t!]
\caption{Efficiencies $\varepsilon_X^{\ell \ell'}$ for the heavy neutrino signals with $\mnmw = (2.5,1)$ TeV in the nine samples corresponding to $\ell,\ell'=e,\mu,\tau$ (rows) and the twelve different same-sign and opposite-sign dilepton final states $X$ (columns) separated according to the `low' and `high' $E_T^\text{miss}$ selection criterium (see text).
\label{tab:eff1}}
\begin{center}
\begin{tabular}{ccccccccccccc}
\hline
\hline
& \multicolumn{6}{c}{Same sign} & \multicolumn{6}{c}{Opposite sign} \\
& \multicolumn{3}{c}{Low $E_T^\text{miss}$} & \multicolumn{3}{c}{High $E_T^\text{miss}$}
& \multicolumn{3}{c}{Low $E_T^\text{miss}$} & \multicolumn{3}{c}{High $E_T^\text{miss}$} \\
$\ell \ell'$ $\backslash$ $X$ & $ee$ & $e\mu$ & $\mu\mu$ & $ee$ & $e\mu$ & $\mu\mu$ & $ee$ & $e\mu$ & $\mu\mu$ & $ee$ & $e\mu$ & $\mu\mu$ \\
$ee$ & 20.9 & 0 & 0 & 0.4 & 0 & 0
 & 21.5 & 0 & 0 & 0.4 & 0 & 0
\\
$e\mu$ & 0 & 16.5 & 0 & 0 & 1.8 & 0
 & 0 & 15.8 & 0 & 0 & 2.8 & 0
\\
$e\tau$ & 1.6 & 1.4 & 0 & 1.6 & 1.3 & 0
 & 1.5 & 1.3 & 0 & 1.9 & 1.6 & 0
\\
$\mu e$ & 0 & 11.0 & 0 & 0 & 9.4 & 0
 & 0 & 10.8 & 0 & 0 & 9.3 & 0
\\
$\mu \mu$ & 0 & 0 & 9.0 & 0 & 0 & 8.5
 & 0 & 0 & 8.5 & 0 & 0 & 8.8
\\
$\mu \tau$ & 0 & 1.0 & 0.9 & 0 & 1.9 & 1.7
 & 0 & 1.0 & 0.9 & 0 & 2.1 & 1.8
\\
$\tau e$ &  0.5 & 0.5 & 0 & 3.5 & 3.2 & 0
 & 0.8 & 0.7 & 0 & 3.7 & 3.3 & 0
\\
$\tau \mu$ & 0 & 0.4 & 0.5 & 0 & 2.9 & 2.8
 & 0 & 0.7 & 0.6 & 0 & 3.3 & 3.0
\\
$\tau \tau$ & 0.1 & 0.2 & 0.1 & 0.5 & 0.8 & 0.4
 & 0.2 & 0.2 & 0.1 & 0.5 & 0.9 & 0.4
\\
\hline
\hline
\end{tabular}
\end{center}
\end{table}

\begin{table}[htb]
\caption{The same as in Table~\ref{tab:eff1} for $\mnmw = (2,1)$ TeV.
\label{tab:eff2}}
\begin{center}
\begin{tabular}{ccccccccccccc}
\hline
\hline
& \multicolumn{6}{c}{Same sign} & \multicolumn{6}{c}{Opposite sign} \\
& \multicolumn{3}{c}{Low $E_T^\text{miss}$} & \multicolumn{3}{c}{High $E_T^\text{miss}$}
& \multicolumn{3}{c}{Low $E_T^\text{miss}$} & \multicolumn{3}{c}{High $E_T^\text{miss}$} \\
$\ell \ell'$ $\backslash$ $X$ & $ee$ & $e\mu$ & $\mu\mu$ & $ee$ & $e\mu$ & $\mu\mu$ & $ee$ & $e\mu$ & $\mu\mu$ & $ee$ & $e\mu$ & $\mu\mu$ \\
$ee$ & 21.8 & 0 & 0 & 0.2 & 0 & 0
 & 21.8 & 0 & 0 & 0.3 & 0 & 0
\\
$e\mu$ & 0 & 18.1 & 0 & 0 & 1.3 & 0
 & 0 & 17.3 & 0 & 0 & 1.7 & 0
\\
$e\tau$ & 1.8 & 1.6 & 0 & 1.3 & 1.1 & 0
 & 1.7 & 1.5 & 0 & 1.6 & 1.4 & 0
\\
$\mu e$ & 0 & 15.3 & 0 & 0 & 5.4 & 0
 & 0 & 14.9 & 0 & 0 & 5.4 & 0
\\
$\mu \mu$ & 0 & 0 & 12.7 & 0 & 0 & 5.3
 & 0 & 0 & 12.4 & 0 & 0 & 5.6
\\
$\mu \tau$ & 0 & 1.4 & 1.2 & 0 & 1.6 & 1.4
 & 0 & 1.3 & 1.1 & 0 & 1.7 & 1.5
\\
$\tau e$ &  0.8 & 0.8 & 0 & 2.9 & 2.6 & 0
 & 1.2 & 1.0 & 0 & 3.2 & 2.8 & 0
\\
$\tau \mu$ & 0 & 0.7 & 0.7 & 0 & 2.6 & 2.5
 & 0 & 1.0 & 1.0 & 0 & 2.9 & 2.6
\\
$\tau \tau$ & 0.1 & 0.2 & 0.1 & 0.3 & 0.6 & 0.3
 & 0.2 & 0.3 & 0.2 & 0.4 & 0.7 & 0.3
\\
\hline
\hline
\end{tabular}
\end{center}
\end{table}

We investigate the prospects for measuring the heavy neutrino couplings by simulating pseudo-experiments with selected mixing benchmark points given in Table~\ref{tab:mix}, and assuming that the number of observed events in the pseudo-experiment equals the signal $S_X$ (which differs from one mixing benchmark point to other) plus the SM background $B_X$.
\begin{table}[htb]
\caption{Mixing benchmark points considered in the pseudo-experiments.
\label{tab:mix}}
\begin{center}
\begin{tabular}{cccc}
\hline
\hline
Label &\quad $V_{eN}^R$ & $\quad V_{\mu N}^R$ &\quad $V_{\tau N}^R$ \smallskip \\ \hline
A & 0 & 1 & 0 \\
B & 0.3 & 0.95 & 0 \\
C & 0.3 & 0.7 & 0.65 \\
D & 0.5 & 0.5 & 0.71 \\
E & 0.7 & 0.3 & 0.65 \\
F & 0.95 & 0.3 & 0 \\
G & 1 & 0 & 0 \\
\hline
\hline
\end{tabular}
\end{center}
\end{table}
\begin{figure}[t!]
\begin{center}
\begin{tabular}{c}
\hspace{-0.3cm}\epsfig{file=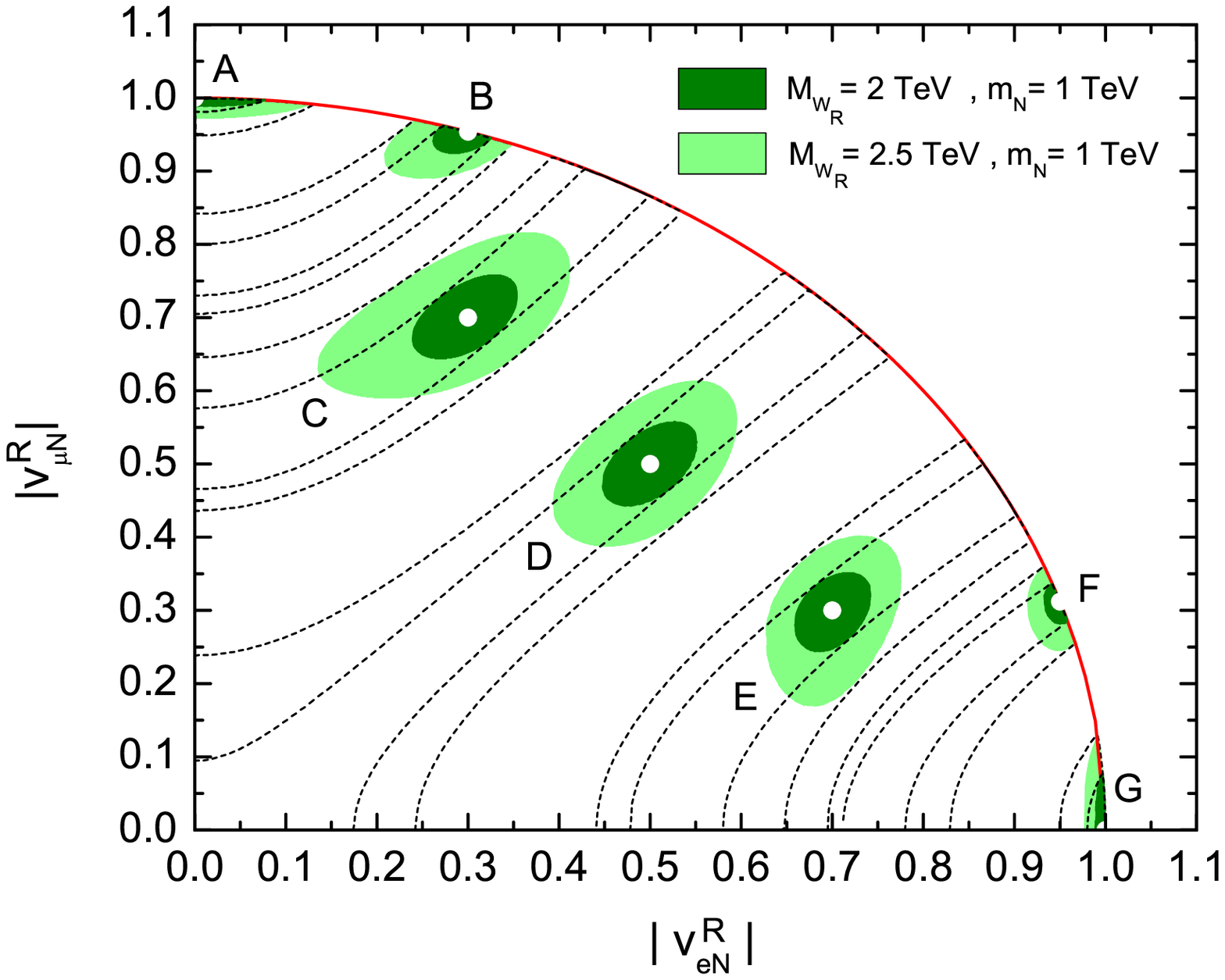,width=8.0cm,clip=} \\ 
\hspace{-0.3cm}\epsfig{file=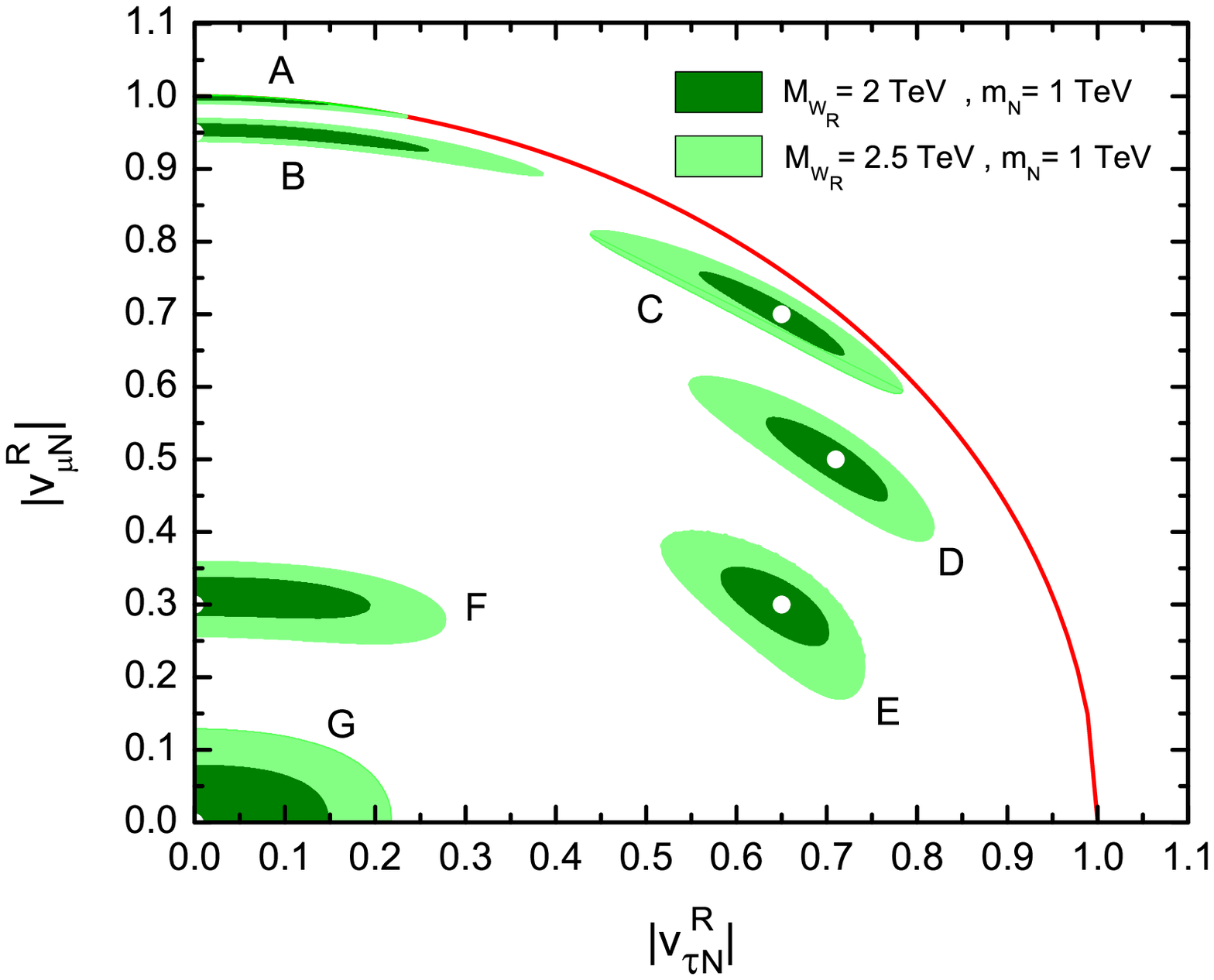,width=8.1cm,clip=} \\
\end{tabular}
\caption{Limits on unitary heavy neutrino mixings ($\kappa=1$) at 68.3~$\%$ CL for the two mass benchmark points $\mnmw = (2.5,1)$~TeV (light green) and $\mnmw = (2,1)$~TeV (dark green) in the $(v_{eN}^R,v_{\mu N}^R)$ plane (top) and in the $(v_{\tau N}^R,v_{\mu N}^R)$ plane (bottom). In the top panel, the dashed lines delimit the 68.3~$\%$ CL regions obtained without separation of the dilepton samples into `high' and `low' $E_T^\text{miss}$.}
\label{fig:lim1}
\end{center}
\end{figure}
\begin{figure}[t!]
\begin{center}
\begin{tabular}{c}
\hspace{-1cm}\epsfig{file=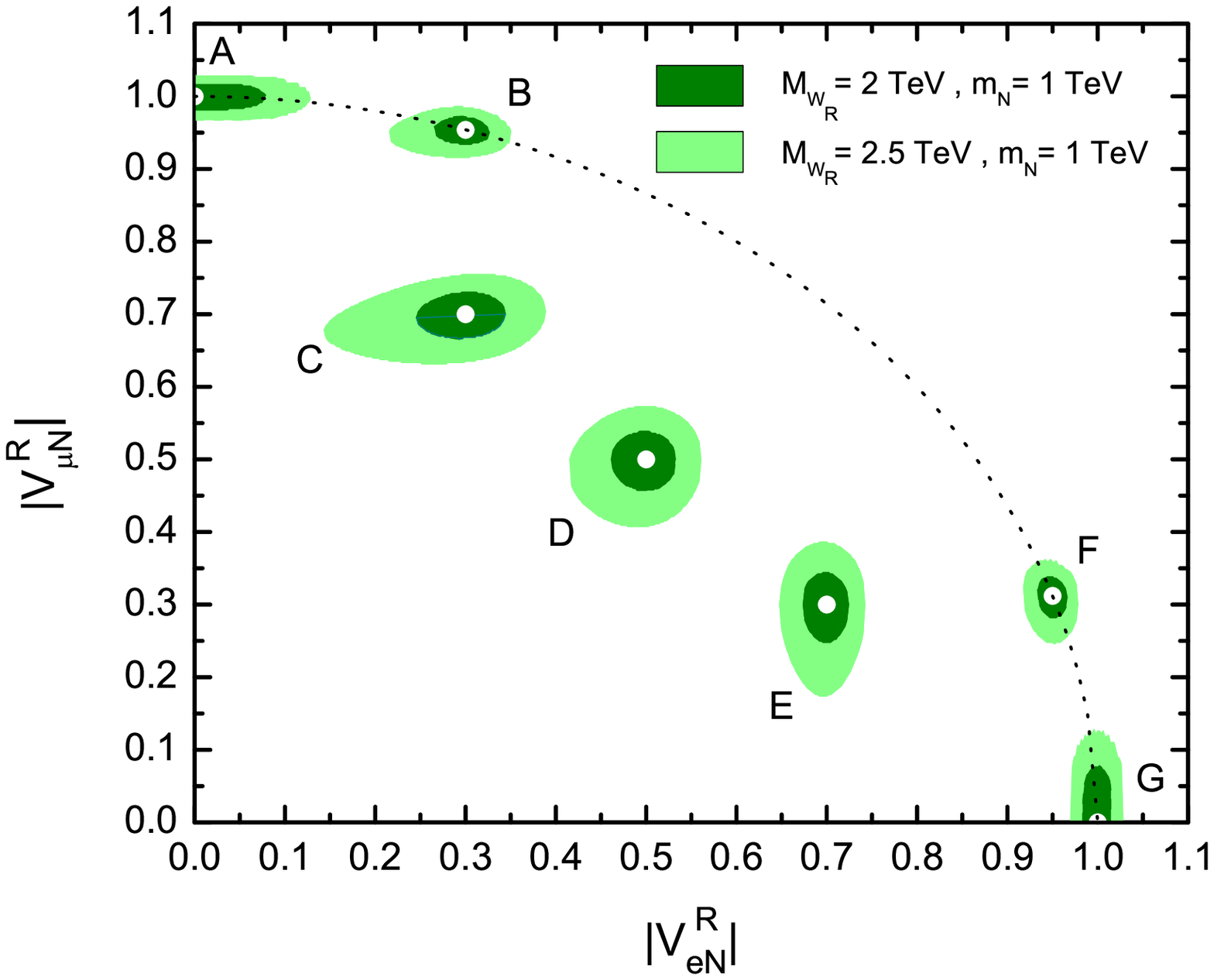,width=7.8cm,height=6cm,clip=} \\ 
\hspace{-1cm}\epsfig{file=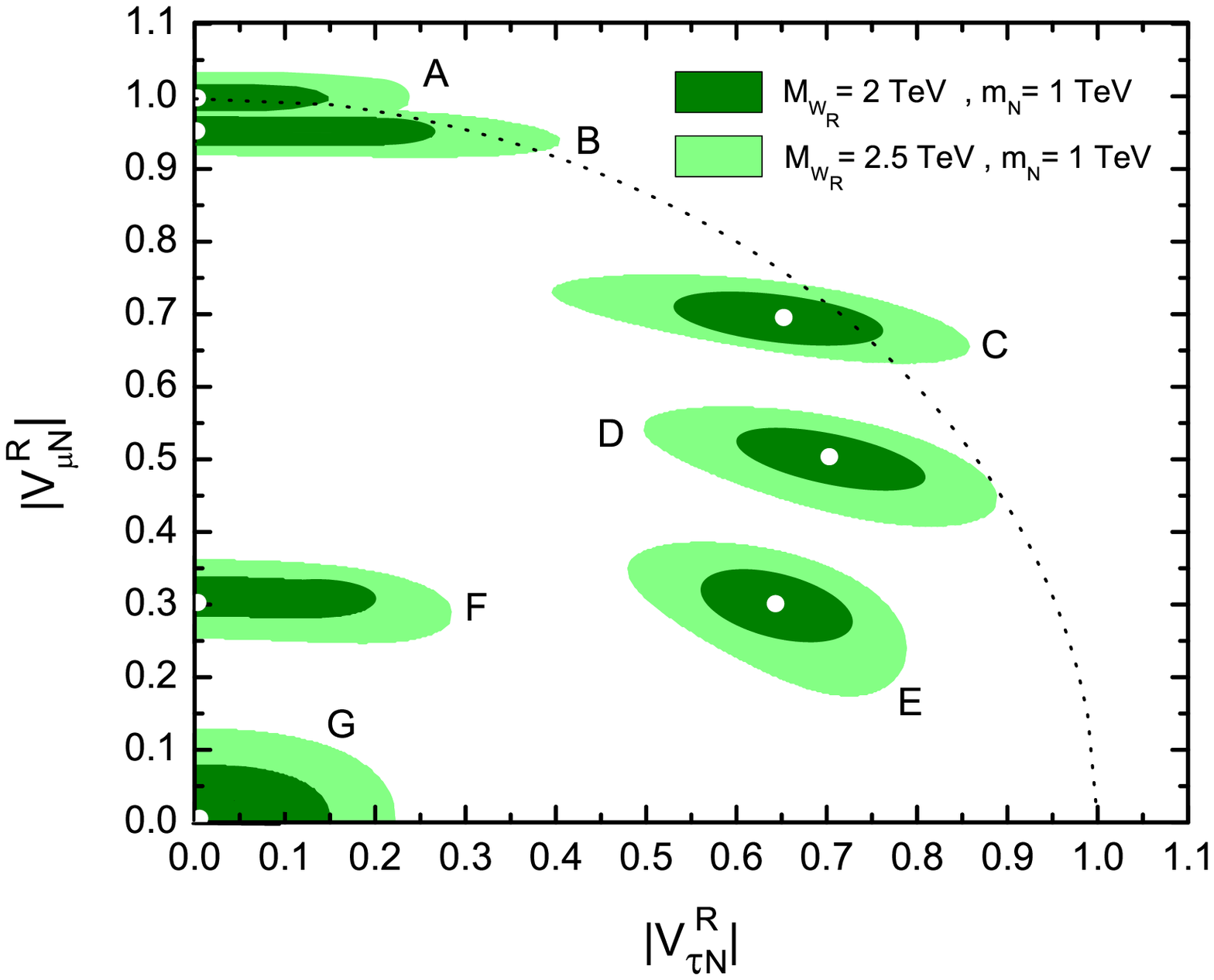,width=7.9cm,height=5.8cm,clip=} \\
\epsfig{file=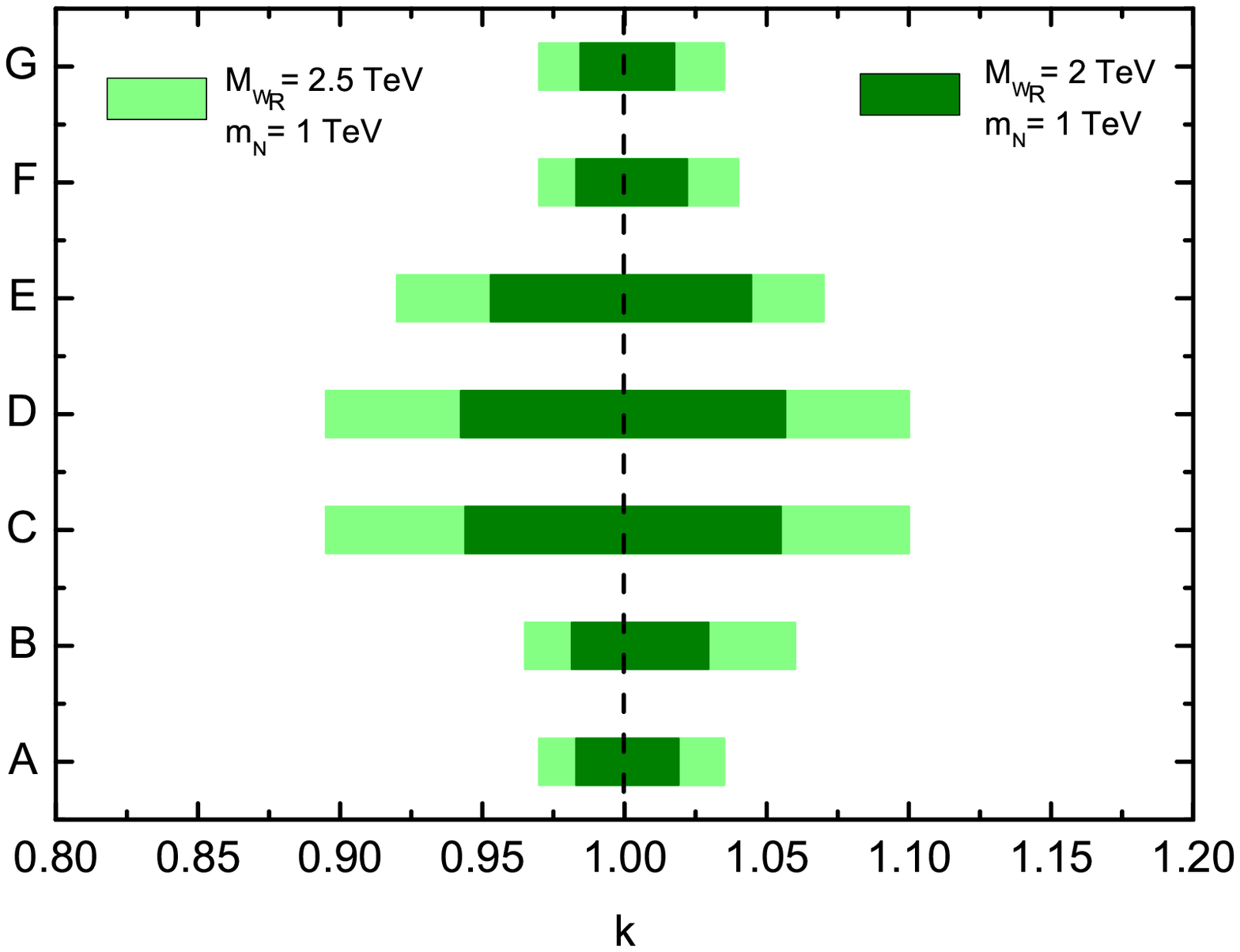,width=7.1cm,clip=}
\end{tabular}
\caption{Limits on the heavy neutrino mixings $V_{\ell N}^R$ at 68.3~$\%$ CL for the two mass benchmark points $\mnmw = (2.5,1)$~TeV (light green) and $\mnmw = (2,1)$~TeV (dark green) in the $(V_{eN}^R,V_{\mu N}^R)$ plane (top) and in the $(V_{\tau N}^R,V_{\mu N}^R)$ plane (middle). In the lower panel, the limits on the normalization parameter $\kappa$ of Eq.~(\ref{ec:kappa}) are shown.}
\label{fig:lim2}
\end{center}
\end{figure}
Then, we extract the signal cross section and the heavy neutrino mixing parameters from the pseudo-data by minimizing the quantity
\begin{equation}
\chi^2 = \sum_X \frac{(N_X^\text{exp}-N_X^\text{th})^2}{N_X^\text{exp}} \,,
\end{equation}
where
\begin{eqnarray}
N_X^\text{exp} = S_X + B_X\,,
\end{eqnarray}
is fixed for each benchmark point, and
\begin{eqnarray}
N_X^\text{th} (\hat \sigma_\text{tot}, \hat v_{eN}^R , \hat v_{\mu N}^R ) & = & \hat \sigma_\text{tot} L \sum_{\ell \ell'} w_{\ell \ell'}  (\hat v_{eN}^R, \hat v_{\mu N}^R) \,\varepsilon_X^{\ell \ell'} \notag \\ & & + B_X \,.
\end{eqnarray}
The above quantity is a function of the two independent flavour mixing parameters $\hat v_{eN}^R$, $\hat v_{\mu N}^R$ and the total cross section $\hat \sigma_\text{tot}$ (we use the hat notation to distinguish the fitted values from the `true' ones used to generate the pseudo-data sample). The inclusion of systematic uncertainties is not essential for our analysis, which is more focused on the method to extract heavy neutrino mixings from data, rather than on the precise value of the limits obtained. In any case, the statistics are moderate for the two mass points considered and, therefore, we do not expect a dramatic change of the results presented here when systematic uncertainties are properly accounted for.

In Fig.~\ref{fig:lim1} we present our results for the two mass benchmark points considered.
The upper plot shows the limits on the two independent normalized mixings $v_{e N}^R$ and $v_{\mu N}^R$. The white dots correspond to the input values used in the pseudo-experiments (see Table~\ref{tab:mix}) and the solid regions are the 68.3\% confidence level (CL) limits on the two independent normalized mixings $v_{e N}^R$ and $v_{\mu N}^R$ extracted from the pseudo-experiments, as discussed above. Notice that the normalized mixings satisfy $|v_{e N}^R|^2 +  |v_{\mu N}^R|^2 + |v_{\tau N}^R|^2 = 1$ by definition, so the region outside the {\em unitary} red arc is not allowed.  The dashed lines correspond to the 68.3\% CL limits that would be obtained without splitting the dilepton samples into `low' and `high' $\etmiss$ (for each mixing benchmark point, the pair of outer lines corresponds to $M_{W_R} = 2.5$ TeV and the inner lines to $M_{W_R} = 2$~TeV). In most cases, the degeneracy that we have discussed qualitatively in Eqs.~(\ref{ec:deg}) is manifest. The only exception occur for points A (G) where the absence of electrons (muons) in the signal implies a mixing with the muon (electron) close to maximal. The lower plot presents the same benchmark points and resulting limits but viewed in the $(v_{\tau N}^R, v_{\mu N}^R)$ plane.

 In Fig.~\ref{fig:lim2} we repeat the analysis for the same benchmark points assuming that the measured value of $\sigma_{W_R}$ corresponds to $g_R = g_L$ and equals the theoretical value presented before for each mass point. The absolute mixings $V_{\ell N}$ equal the normalized ones (in Fig.~\ref{fig:lim1}) multiplied by $\kappa$, which is obtained from the fitted cross section $\hat \sigma_\text{tot}$ and Eq.~(\ref{ec:kappa2}). As in the previous case, the allowed regions for the mixings at 68.3$\%$ CL are shown in the $(V_{e N}^R,V_{\mu N}^R)$ and $(V_{\tau N}^R,V_{\mu N}^R)$ planes (top and middle plots, respectively). The corresponding limits for the parameter $\kappa$ are given in the lower panel (color codes are the same as in Fig.~\ref{fig:lim1}). We observe that, provided the statistics are good and the systematic uncertainties for the total cross sections (which we have ignored) are controlled at the $\sim 10\%$ level or better, the unitarity of the mixings can also be tested.

\section{Discussion}

In case new physics is discovered at the LHC, it will be essential to determine its properties.
Here, we have addressed one particular example that represents a variety of new physics models related to neutrino mass generation: the production of heavy neutrinos in the context of a left-right symmetric model. We have investigated the determination of the couplings of a heavy neutrino $N$ to the charged leptons, showing that an inclusive event counting is not sufficient for this purpose. Basically, this happens because the electrons and muons in the final state can be produced promptly in the $N$ decays, or from secondary decays of $\tau$ leptons produced in the decays of the heavy neutrinos. We have pointed out that one possible solution for this degeneracy is to discriminate the dilepton events based on the missing energy, so that prompt electrons and muons can be distinguished from the secondary ones.

Obviously, this kind of degeneracy is not a specific feature of this model, but a generic one in any SM extension with new particles coupling to the charged leptons. The most immediate example is type-I seesaw without left-right symmetry, in
 \begin{equation}
pp \to W^* \to \ell N \to \ell \ell' jj \,,
\end{equation}
although in this case observing the signal is very demanding~\cite{delAguila:2007em}, let alone measuring the heavy neutrino couplings. More interesting is the production of heavy neutrino pairs via a new $Z'$ boson~\cite{delAguila:2007ua,Huitu:2008gf,AguilarSaavedra:2009ik,Li:2010rb},
 \begin{equation}
pp \to Z' \to N N \to \ell \ell' jjjj \,.
\end{equation}
Similarly, in type-III seesaw one has dilepton signals resulting from the production of the new heavy leptons $E,N$
 \begin{equation}
pp \to E N \to \ell \ell' jjjj \,.
\end{equation}
In all these cases, the procedure presented here can be straightforwardly applied to determine the couplings of the heavy leptons. In case more than one heavy neutrino is produced at the LHC, the method presented here can be generalized by introducing appropriate kinematical cuts to single out the contribution of each heavy neutrino. The practical implementation of this sophistication (which would be necessary if several heavy neutrinos were discovered) strongly depends on the particular values of the heavy neutrino masses, and falls beyond the scope of this work.

\acknowledgements
This work has been supported by the and \textit{Funda\c c\~ao para a Ci\^encia e a Tecnologia} (FCT, Portugal) and \textit{Ministerio de Ciencia e Innovaci\'on} (MICINN, Spain) under the bilateral project ``Signals of new fermions at colliders'' (FCT/1683/27/1/2012/S, AIC-D-2011-0811). J.A.A.S. acknowledges support from the projects FPA2006-05294, FPA2010-17915 and the Junta de Andaluc\'{\i}a projects FQM 101, FQM 03048 and FQM 6552. F.R.J. thanks the CERN Theory Division for hospitality during the final stage of this work and acknowledges support from the EU Network grant UNILHC PITN-GA-2009-237920 and from the FCT projects CERN/FP/123580/2011, PTDC/FIS/098188/2008 and CFTP-FCT UNIT 777.

\end{document}